\documentclass[
 amsmath,amssymb,
pra, twocolumn
]{revtex4-2}
\usepackage{amsmath}
\usepackage{physics}
\usepackage{graphicx}
\usepackage{dcolumn}
\usepackage{bm}
\usepackage{xcolor}

\definecolor{themeColour1}{RGB}{255,185,72} 
\definecolor{themeColour2}{RGB}{16,130,46} 
\definecolor{themeColour3}{RGB}{23, 113, 191} 
\definecolor{themeColour4}{RGB}{149,18,114} 

\usepackage[colorlinks,pdfusetitle,urlcolor=blue,citecolor=blue,linkcolor=blue,bookmarksnumbered,plainpages=false]{hyperref}

\newcommand{\rA}{\mathbf{r}_\text{A}}
\newcommand{\rD}{\mathbf{r}_\text{D}}
\newcommand{\dD}{\mathbf{d}_\text{D}}
\newcommand{\dA}{\mathbf{d}_\text{A}}
\newcommand{\rOpt}{\mathbf{r}''}
\newcommand{\G}{\mathbb{G}}
\newcommand{\eps}{\bar{\varepsilon}}
\renewcommand{\i}{\mathrm{i}}
\newcommand{\merit}{R}
\newcommand{\fancyRe}{\mathfrak{Re}}
\newcommand{\fancyIm}{\mathfrak{Im}}
\newcommand{\new}[1]{#1}
\newcommand{\secondNew}[1]{#1}

\begin{document}

\title{Towards nanophotonic optical isolation via inverse design of energy transfer in nonreciprocal media}

\author{Claire M. Cisowski}
 \email{clairemarie.cisowski@glasgow.ac.uk}
\author{Madeline C. Waller}%
\author{Robert Bennett}%
\affiliation{%
 School of Physics and Astronomy, University of Glasgow, G12 8QQ, Glasgow, Scotland, UK
}%

\date{\today}

\begin{abstract}
In this work we generalise the adjoint method of inverse design to nonreciprocal media. As a test case, we use three-dimensional topology optimization via the level-set method to optimise one-way energy transfer for point-like source and observation points. To achieve this we introduce a suite of tools, chiefly what we term the `Faraday-adjoint' method which allows for efficient shape optimization in the presence of magneto-optical media. We carry out an optimization based on a very general equation that we derive for energy transfer in a nonreciprocal medium, and link finite-different time-domain numerics to analytics via a modified Born series generalised to a tensor permittivity. This work represents a stepping stone towards practical nanophotonic optical isolation, often regarded as the `holy grail' of integrated photonics.  
\end{abstract}

\maketitle

\section{Introduction}

Reciprocity defines much of our experience of everyday life. Consider, for instance, the simple fact that when we can hear someone, we can be certain that they can hear us as well. Similar ideas apply to light --- if a laser beam transmits a message, we would expect that the same information would be transferred if the source and observer were to swap positions. This symmetry can be broken if the intervening medium is nonreciprocal, which can be taken advantage of in the construction of critical technological devices such as optical isolators and circulators (see \cite{Caloz2018} for a comprehensive review). Such components are vital whenever one-way propagation is needed, so find uses across communication technologies in, for example, eliminating unwanted back-reflections (see, e.g., \cite{jalasWhatWhatNot2013}). 

\begin{figure}
\includegraphics[]{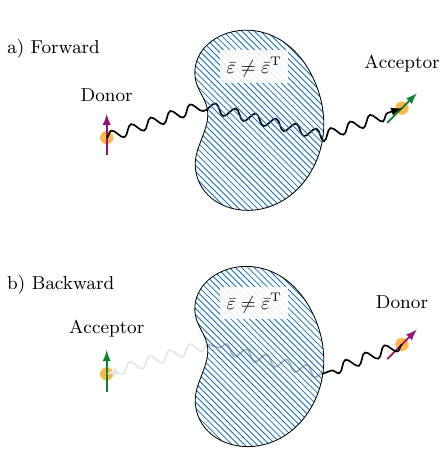}
\caption{The general idea of optical isolation,  illustrated in terms of the system we will consider. Both panels show a schematic of two atoms or molecules (modelled here as point-like dipoles) in the presence of an arbitrarily-shaped nonreciprocal environment of asymmetric tensor permittivity $\eps\neq \eps^\mathrm{T}$. When the donor and acceptor positions are reversed as shown in b), the rate of energy transfer will not in general be the same due to the nonreciprocal nature of the medium. Isolation (simultaneous increase of the `forward' rate and decrease of the `backward' rate) via topological optimization of the intervening structure is the overall goal pursued in this work.}\label{generalRETFig}
\end{figure}

There are a variety of routes to non-reciprocity. By far the most common method uses the Faraday effect \cite{faradayFaradayDiaryBeing1932}, where the non-reciprocity comes from the interplay between material response and an applied external magnetic field $\mathbf{B}_0$. No matter whether a beam propagates parallel or anti-parallel to $\mathbf{B}_0$, the Faraday effect causes the polarization of an incoming beam to be rotated in the \emph{same} direction (in the lab frame), so that back-reflected light can be filtered out. At the cm-scale, technologies based on the Faraday effect are very well-developed, with nonreciprocal systems being discussed in the literature as early as the second half of the 19th century \cite{lordrayleighMagneticRotationLight1901}. The isolator has, however, been particularly resistant to modern goals of miniaturization and integration into photonic systems. This is chiefly due to the difficulties in integrating traditional Faraday media (rare earth iron garnets) into silicon-based platforms due to the lossy and CMOS-incompatible nature of magneto-optical materials (see, e.g.,~\cite{shojiMagnetoopticalIsolatorSilicon2008,Bi2011,tienSiliconRingIsolators2011a,sobuGaInAsPInPMZI2013,Huang2016,Zhang:19}, and for a recent review see \cite{srinivasanReviewIntegratedMagnetooptical2022}). A class of `magnetless' devices based on the inherent non-linearity of certain integrated waveguides has shown some promise (e.g.~\cite{galloAllopticalDiodePeriodically2001,soljacicNonlinearPhotonicCrystal2003,fanAllSiliconPassiveOptical2012,Chang2014,Mahmoud2015,DelBino:18}) but have been shown to have intrinsic limitations relating to back-propagation of noise for higher input powers \cite{shiLimitationsNonlinearOptical2015a}. Other approaches based on spatiotemporal modulation \cite{doerrOpticalIsolatorUsing2011,PhysRevLett.109.033901,doerrSiliconPhotonicsBroadband2014,Sounas2017,Kittlaus2018} or optomechanical coupling \cite{Sohn2018} have also demonstrated encouraging results, but the former are specific to continuous-wave operation and all can cause undesirable frequency components to become populated. Aside from having a strong non-reciprocity, any feasible integrated optical isolator must simultaneously satisfy (at least) bandwidth, input power, linearity and CMOS compatibility constraints. 

Silicon's low loss and the existence of vast CMOS infrastructure suggest that it may yet be the material of choice for the construction of a passive, linear, integrated optical isolator. It is often overlooked that silicon itself exhibits a small Faraday rotation of around $15^\circ$cm$^{-1}$T$^{-1}$ at a wavelength of $1550$nm, two orders of magnitude lower than rare earth iron garnets (see, e.g., \cite{srinivasanReviewIntegratedMagnetooptical2022}).  Constructing an integrated isolator entirely out of silicon is, at first sight, unrealistic due to the cm-scale propagation distances required to obtain the required 45$^\circ$ polarization rotation. One creative approach to solving this problem has been taking advantage of silicon's very weak Faraday effect by `wrapping' silicon waveguides in order to gain a sufficient propagation length in a small enough footprint \cite{jalasFaradayRotationSilicon2017}. Aside from the general requirements on an isolator discussed above, there are a variety of stringent design constraints on such a device, for example requiring that any bend in the waveguide must possess the right birefringence to rotate the polarization by the angle of the bend itself (amongst other criteria) \cite{jalasFaradayRotationSilicon2017}.

Inspired by the above, we will introduce a new method of designing the nonreciprocal response of a device. This is based on \emph{inverse design}, where improved structures are discovered by an efficient algorithm, as opposed to a traditional `intuition based' approach. Instead of a designer specifying a structure and then testing it against a set of desired characteristics, inverse design allows the designer to specify only the goal (and any applicable constraints), allowing an efficient free-form algorithm to build the structure. The advantage in this lies with the fact that  several interdependent characteristics can be simultaneously optimized, and the resulting structures do not rely on the intuition or previous experience of the designer.  

Inverse design has its origins in decades-old mechanical problems \cite{topology_meca_BENDSOE1988} but has undergone an explosion of interest in recent years due to computational advances, now being one of the standard tools available in nanophotonics (for a review, see \cite{Molesky2018}). It has been used to design and optimise photonic crystals ~\cite{Borel:04,Burger2004InversePT}, waveguides, \cite{Jensen2004}, routers \cite{yangInversedesignedNonreciprocalPulse2020}, resonators \cite{luInverseDesignThreedimensional2011,yuGeneticallyOptimizedOnchip2017,linTopologyOptimizationMultitrack2017,ahnPhotonicInverseDesign2022}, plasmonic devices \cite{duhringOptimizationExtraordinaryOptical2013,zengInverseDesignPlasmonic2021} metasurfaces \cite{shenUltrahighefficiencyMetamaterialPolarizer2014,Pestourie18,Li2022} demultiplexers \cite{Piggott2015InverseDA,frellsenTopologyOptimizedMode2016} and even optical lattice patterns \cite{kouznetsovInverseDesignAssisted2022,Romekpaper}. Outside nanophotonics, the same formalism finds applications diverse fields such as microfluidics \cite{Borrvall2003}, antenna design~\cite{Kiziltas2003,Erentok2011} and phononic band gap optimization~\cite{Bonnecaze2003}, amongst others. Inverse-designed devices have been shown in various contexts to have performance vastly in excess of any traditionally-designed system (see, e.g., \cite{wangHighperformanceSlowLight2012,deformationpaper}). This means, for example, that the complex structures required in the silicon-based isolator of \cite{jalasFaradayRotationSilicon2017} would instead emerge `naturally' in the inverse design process, and its performance may be dramatically enhanced --- perhaps enough to be technologically relevant. 

There is a reason, however, why inverse design has not yet been applied to devices reliant on nonreciprocal media (the inverse design for the nonreciprocal router proposed in \cite{yangInversedesignedNonreciprocalPulse2020} was applied to reciprocal media to enhance coupling between various waveguides, and \cite{kielEnhancedFaradayRotation2021} was based on pure Bayesian shape optimization, without taking the physical properties of nonreciprocal media into account). The problem lies with the fact that inverse design in its modern form is made possible by efficient computational techniques that allow for simultaneous optimization of essentially arbitrary numbers of degrees of freedom. Arguably the most prominent technique is the \emph{adjoint method} \cite{pironneauOptimumDesignFluid1974,jamesonAerodynamicDesignControl1988,gilesIntroductionAdjointApproach2000}, which reduces the number of simulations required to optimise over $N$ parameters down to just \emph{two} (independent of $N$). These two simulations are the `forward' one (where, in the context of electromagnetism, the radiation propagates from the physical source to the observer) and the `adjoint' one (where source and observer are swapped). The way radiation propagates through the \emph{same} system but in the opposite direction is then used to determine a gradient in the optimization space. It is clear that in a nonreciprocal system the adjoint method will not apply in the same fashion as in a reciprocal one. In particular, the adjoint simulation in a nonreciprocal medium will be physically distinct from the forward one, meaning the two simulations undertaken in the adjoint method do not represent the same physical system. We will show how this difficulty can be elegantly sidestepped using the Green's tensor based inverse design approach introduced in \cite{QEDBennett_2020} to carry out three-dimensional topology optimization of nonreciprocal media using a modified adjoint method. We will use resonant energy transfer (RET) as illustrated in Fig.~\ref{generalRETFig} as a simple test observable to demonstrate the viability of the magneto-optical inverse design tools we are introducing, paving the way towards highly-optimised all-silicon optical isolators for photonics applications, or indeed to optimization of geometries for traditional rare-earth iron garnet based components.

This paper is organised as follows. In section \ref{RETinNRMedia} we derive an equation describing the rate of energy transfer in nonreciprocal media. We then validate this model against finite-difference time-domain calculations in \ref{weakNonRecipSection}, simultaneously introducing another tool for semi-analytically calculating the nonreciprocal response of an arbitrarily-shaped medium. In section \ref{inverseDesignGExplanationSection} we provide the formalism for adapting the Green's tensor-based adjoint method to nonreciprocal media, and finally in Section \ref{demonstrationSection} we carry out three-dimensional level-set optimization of energy transfer isolation to provide a proof-of-principle for the methods and techniques introduced.   


\section{RET in nonreciprocal media}\label{RETinNRMedia}

    Resonance energy transfer is a ubiquitous process across the sciences, enabling transport of energy in plants and having applications in, for example, artificial photonic complexes \cite{mohapatraForsterResonanceEnergy2018}. We will use resonance energy transfer from a point-like donor atom or molecule and to point-like acceptor atom or molecule as a test observable for our optimization of nonreciprocal media, essentially because it is the simplest two-center process that one can envisage. The rate of energy transfer is well-known in a wide variety of situations, ranging from the simple electrostatic treatments of F\"{o}rster \cite{forsterZwischenmolekulareEnergiewanderungUnd1948} to the generalised QED-based theory in vacuum \cite{andrewsUnifiedTheoryRadiative1989, andrewsVirtualPhotonsDipole2004} and in the presence of dispersive and absorbing media \cite{dungIntermolecularEnergyTransfer2002}. To the best of our knowledge, the rate of energy transfer has not been calculated in nonreciprocal media. We therefore present an original calculation of such a rate here.

We set up a system of a donor and acceptor, where energy from the donor is released and transferred to the acceptor, through a possibly nonreciprocal medium. The Hamiltonian for this system is written as;
\begin{equation}
    H = H_0 + H^{\mathrm{A}}_{\mathrm{int}} + H^{\mathrm{D}}_{\mathrm{int}},
\end{equation}
where
\begin{equation}
    H_0 = H_{\mathrm{rad}} + H^{\mathrm{A}}_{\mathrm{mol}} + H^{\mathrm{D}}_{\mathrm{mol}},
\end{equation}
and $H_{\mathrm{rad}}$ is the Hamiltonian of the radiation field, $H^\xi_\mathrm{mol}$ is the Hamiltonian of the atom or molecule $\xi$ for which we assume that the eigenstates are known, and
\begin{equation}
    H^\alpha_{\mathrm{int}}=-\mathbf{\hat{d}}_\alpha \cdot \mathbf{\hat{E}}(\mathbf{r}_\alpha),
    \label{eq:InteractionHamiltonian}
\end{equation}
where  $\hat{\mathbf{d}}_\alpha$ is the transition dipole moment operator of molecule $\alpha$, and $\hat{\mathbf{E}}(\mathbf{r}_\alpha)$ is the quantised electric field at the position, $\mathbf{r}_\alpha$, of the molecule $\alpha$. For energy transfer the initial and chosen final states of the system are:
\begin{equation}
    \ket{i}=\ket{e_{\mathrm{D}},g_{\mathrm{A}};0}, \qquad
    \ket{f}=\ket{g_{\mathrm{D}},e_{\mathrm{A}};0},
\end{equation}
where $g_{\mathrm{D}}$ $(g_{\mathrm{A}})$ denotes the ground state of the donor (acceptor), $e_{\mathrm{D}}$ $(e_{\mathrm{A}})$ the excited state of the donor (acceptor) and $0$ is the ground state of the electromagnetic field. 

We use macroscopic QED \cite{grunerCorrelationRadiationfieldGroundstate1995,dungThreedimensionalQuantizationElectromagnetic1998} to describe the electric field. This very general theory allows the effect of an environment near the donor and acceptor to be taken into account. Ordinarily, this environment would be described using a scalar position- and frequency-dependent permittivity $\varepsilon(\mathbf{r},\omega)$ (we will only consider materials of unit relative permeability here). A nonreciprocal medium, however, has a permittivity tensor $\eps(\mathbf{r},\omega)$. A medium is nonreciprocal if $\eps(\mathbf{r},\omega) \neq \eps^\mathrm{T}(\mathbf{r},\omega)$, where T denotes the transpose. 

The expression for the quantised electric field takes the following form in the presence of nonreciprocal media \cite{buhmannMacroscopicQuantumElectrodynamics2012};
\begin{align}
    \mathbf{E}(\mathbf{r})= \int_0^{\infty} d\omega \int d^3 \mathbf{s} \, \mathbf{F}(\mathbf{r},\mathbf{s},\omega) \cdot \mathbf{\hat{f}} (\mathbf{s},\omega) + \text{H.c.} \label{eq:Efield}
\end{align}
where $\hat{\mathbf{f}}_{\lambda}(\mathbf{r}',\omega)$ is an annihilation operator for a polaritonic excitation at position $\mathbf{r}'$ and with frequency $\omega$, and its Hermitian conjugate is the corresponding creation operator. These operators obey bosonic commutation relations;
\begin{align}
    \left[ \mathbf{\hat{f}}(\mathbf{r},\omega),\mathbf{\hat{f}}(\mathbf{r}',\omega')\right]&=\left[ \mathbf{\hat{f}}^\dagger(\mathbf{r},\omega),\mathbf{\hat{f}}^\dagger(\mathbf{r}',\omega')\right]=0 
    \\ \left[ \mathbf{\hat{f}}(\mathbf{r},\omega),\mathbf{\hat{f}}^\dagger(\mathbf{r}',\omega')\right] &= \bm{\delta}(\mathbf{r}-\mathbf{r}') \delta(\omega-\omega').
\end{align}
where $\bm{\delta}(\mathbf{r}-\mathbf{r}')=\mathrm{diag}(1,1,1)\delta(\mathbf{r}-\mathbf{r}')$.
$\mathbf{F}(\mathbf{r},\mathbf{s},\omega)$ is a function we have defined as;
\begin{align}
    \mathbf{F}(\mathbf{r},\mathbf{s},\omega)&= \i\mu_0 \sqrt{\frac{\hbar}{\pi}} \omega^{3/2}\int d^3\mathbf{r}' \, \G(\mathbf{r},\mathbf{r}',\omega) \cdot \mathbf{R}(\mathbf{r}',\mathbf{s},\omega) 
\end{align}
where $\G(\mathbf{r},\mathbf{r}',\omega)$ is the Green's tensor which obeys the following generalised Helmholtz equation \cite{buhmannMacroscopicQuantumElectrodynamics2012};
\begin{align}\label{generalisedHelmholtz}
    \delta(\mathbf{r}-\mathbf{r}')=&\left[\nabla \times \nabla \times -\frac{\omega^2}{c^2} \right] \G(\mathbf{r},\mathbf{r}'\omega) \notag \\ 
    &-i\mu_0\omega \int d^3\mathbf{s} \mathbf{Q}(\mathbf{r},\mathbf{s},\omega) \cdot \G(\mathbf{s},\mathbf{r}',\omega).
\end{align}
where $\mathbf{R}$ is a square root of the positive definite tensor field $\fancyRe[\mathbf{Q}]$;
\begin{equation}
    \int d^3 \mathbf{r}'' \mathbf{R}(\mathbf{r},\mathbf{r}'',\omega) \cdot \mathbf{R}^\dag (\mathbf{r}'',\mathbf{r}',\omega) = \fancyRe[\mathbf{Q}(\mathbf{r},\mathbf{r}',\omega)].
\end{equation}
where $\mathbf{Q}(\mathbf{r},\mathbf{r}',\omega)$ is the conductivity tensor and, following \cite{buhmannMacroscopicQuantumElectrodynamics2012}, we have introduced generalized real and imaginary parts of a tensor field according to;
\begin{align}
    \fancyRe [\mathbf{T}(\mathbf{r},\mathbf{r}')]&=\frac{1}{2} [\mathbf{T}(\mathbf{r},\mathbf{r}')+\mathbf{T}^\dag(\mathbf{r},\mathbf{r}')]
    \label{eq:RealTensor}
    \\
    \fancyIm [\mathbf{T}(\mathbf{r},\mathbf{r}')]&=\frac{1}{2\i} [\mathbf{T}(\mathbf{r},\mathbf{r}')-\mathbf{T}^\dag(\mathbf{r},\mathbf{r}')] 
\end{align}

\begin{figure}
\includegraphics[]{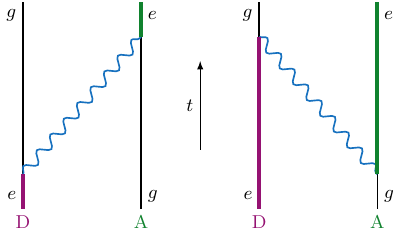}
\caption{The two distinct time-orderings appearing in the matrix element \eqref{eq:MatrixEl}. Time runs from bottom to top, and thick lines represent excited states of the donor D and acceptor A. The resonant contribution comes from the diagram on the left. While both diagrams have off-resonant contributions, these cancel as discussed in detail in the main text.  }\label{timeOrderingsFig}
\end{figure}

The matrix element for two-body resonant energy transfer can be written in the form \cite{PhysRevA.106.043107};
\begin{equation}
    M_{fi} =-\sum_{p} \bra{f} \Biggl[\frac{H^{\mathrm{A}}_{\mathrm{int}} H^{\mathrm{D}}_{\mathrm{int}} }{\hbar cp-E_{\mathrm{eg}}} +\frac{H^{\mathrm{D}}_{\mathrm{int}} H^{\mathrm{A}}_{\mathrm{int}}}{\hbar cp+E_{\mathrm{eg}}} \Biggr] \ket{i} \label{eq:MatrixEl}
\end{equation}
with the two terms representing distinct time-orderings as illustrated in Fig.~\ref{timeOrderingsFig} and the sum running over all possible momenta $p$ of the exchanged photon.
Substituting in our expressions for the interaction Hamiltonian, \eqref{eq:InteractionHamiltonian}, and for the electric field, \eqref{eq:Efield}, and making use of the integral relation applicable in nonreciprocal media \cite{buhmannMacroscopicQuantumElectrodynamics2012};
\begin{align}
    \fancyIm \G(\rA,\rD,\omega) &= \mu_0 \omega \int d^3\mathbf{r}' \int d^3\mathbf{r}'' \notag \\ 
    \G(\rA,\mathbf{r}',\omega)
    \cdot &\fancyRe\mathbf{Q}(\mathbf{r}',\mathbf{r}'',\omega) \cdot \G^{\dag}(\rD,\mathbf{r}'',\omega)
\end{align}
we can rewrite the matrix element \eqref{eq:MatrixEl} as;
\begin{align}
    M_{fi}=&- \frac{\mu_0}{\pi} \int_0^{\infty} d\omega \, \dA^\uparrow \cdot  \frac{\omega^{2} \, \fancyIm \G(\rA,\rD,\omega)}{\omega-\omega_\mathrm{D}} \cdot \dD^\downarrow \notag \\
    &- \frac{\mu_0}{\pi} \int_0^{\infty} d\omega \, \dD^\downarrow \cdot  \frac{\omega^{2} \, \fancyIm \G(\rD,\rA,\omega)}{\omega+\omega_\mathrm{D}} \cdot \dA^\uparrow 
\end{align}
where we have defined
\begin{align}
\hbar cp&=\hbar \omega,  & E_{\mathrm{eg}} &= \hbar \omega_{\mathrm{D}}\\
 \dD^\downarrow  &\equiv \bra{g_\mathrm{D}}\mathbf{\hat{d}}_\mathrm{D}\ket{e_\mathrm{D}},  & \dD^\uparrow  &\equiv \bra{e_\mathrm{D}}\mathbf{\hat{d}}_\mathrm{D}\ket{g_\mathrm{D}},
\\ \dA^\downarrow  &\equiv \bra{g_\mathrm{A}}\mathbf{\hat{d}}_\mathrm{A}\ket{e_\mathrm{A}},  & \dA^\uparrow  &\equiv \bra{e_\mathrm{A}}\mathbf{\hat{d}}_\mathrm{A}\ket{g_\mathrm{A}}.
\end{align}
The frequency integrals have poles on the real axis, so we let the eigenenergies of the atom take on a small imaginary part $\epsilon$. This means that the poles shift to positions $\pm (\omega_\mathrm{D} + i\epsilon)$, allowing the frequency integral to be evaluated by closing the contour in the upper half of the complex plane \cite{jenkinsQuantumPathwaysResonance2004,jonesResonanceEnergyTransfer2019}. We find that;
\begin{align}
   &\lim_{\epsilon \rightarrow 0+} \int^{\infty}_0 d\omega \frac{\omega ^2 \fancyIm \G(\mathbf{r},\mathbf{r}',\omega)}{\omega_{\mathrm{D}} + \omega +i\epsilon} \notag \\ 
    =  &-\frac{1}{2} \int_0^{\infty} d\xi \, \xi^2 \,  \Biggl[ \frac{\G(\rD,\rA,i\xi)}{i\xi + \omega_\mathrm{D}} -  \frac{ \G^\mathrm{T}(\rA,\rD,i\xi)}{i\xi - \omega_\mathrm{D}} \Biggr]
\end{align}
and
\begin{align}
   \lim_{\epsilon \rightarrow 0+} &\int^{\infty}_0 d\omega \frac{\omega ^2 \fancyIm \G(\mathbf{r},\mathbf{r}',\omega)}{\omega_{\mathrm{D}} - \omega +i\epsilon} \notag \\ 
    =  -\frac{1}{2}&\int_0^{\infty} d\xi \, \xi^2 \Biggl[ \frac{\G(\rA,\rD,i\xi)}{i\xi - \omega_\mathrm{D}} -  \frac{\G^\mathrm{T}(\rD,\rA,i\xi)}{i\xi + \omega_\mathrm{D}} \Biggr] \notag \\ 
    &+  \pi \omega_\mathrm{D}^2\G(\rA,\rD,\omega_\mathrm{D})
\end{align}
Summing these two contributions, we find the full matrix element to be;
\begin{align}\label{matrixElementFull}
   & M_{fi} = -\mu_0 \omega_\mathrm{D}^2 \dA^\uparrow \cdot \G(\rA,\rD,\omega_\mathrm{D}) \cdot \dD^\downarrow 
    +\frac{\mu_0}{2\pi} \int_0^{\infty} d\xi \notag \\
    &\times \xi^2 \Biggl( \dA^\uparrow \cdot \biggl[ \frac{\G(\rA,\rD,i\xi)}{i\xi - \omega_\mathrm{D}} - \frac{\G^\mathrm{T}(\rD,\rA,i\xi)}{i\xi + \omega_\mathrm{D}} \biggr]   \cdot \dD^\downarrow \notag \\
    &+ \dD^\downarrow \cdot \biggl[ \frac{\G(\rD,\rA,i\xi)}{i\xi + \omega_\mathrm{D}} - \frac{ \G^\mathrm{T}(\rA,\rD,i\xi)}{i\xi - \omega_\mathrm{D}} \biggr] \cdot \dA^\uparrow \Biggr).
\end{align}
It is easily observed that all terms under the integral vanish (since $\mathbf{a}\cdot \mathbb{B} \cdot \mathbf{c} = \mathbf{c}\cdot \mathbb{B}^\mathrm{T} \cdot \mathbf{a} $ for any vectors $\mathbf{a}$ and $\mathbf{c}$, and matrix $\mathbb{B}$), leaving only the first term:
\begin{equation}
     M_{fi} = -\mu_0 \omega_\mathrm{D}^2 \, \dA^\uparrow \cdot \G(\rA,\rD,\omega_\mathrm{D}) \cdot \dD^\downarrow \label{eq:MfiNRSimp}
\end{equation}
We note that this is the same formula that is obtained for the reciprocal case. Of course, the results will be different as the $\G$ that actually goes into the above equation will be that for nonreciprocal media. 

We now use the matrix element \eqref{eq:MfiNRSimp} in Fermi's Golden Rule;
\begin{align}\label{rateNonRecipAnalytic}
\Gamma &= \frac{2\pi}{\hbar}|M_{fi}|^2 \delta(E_{\mathrm{I}}-E_f) \notag \\
&= \frac{2\pi \mu_0^2 \omega_\mathrm{D}^4}{\hbar} |\dA^\uparrow \cdot \G(\rA,\rD,\omega_\mathrm{D}) \cdot \dD^\downarrow|^2 
\end{align}
where we have additionally assumed real dipole moments. This formula, valid for both reciprocal and nonreciprocal media, will be the basis of the inverse design discussed in the remainder of this article. Before that, there is one more interesting nonreciprocal property of \eqref{rateNonRecipAnalytic} worth mentioning. In reciprocal media, $\G(\mathbf{r},\mathbf{r}',\omega) = \G^\mathrm{T}(\mathbf{r}',\mathbf{r},\omega)$, which implies that 
\begin{equation}\label{reciprocityWithVectors}
    \mathbf{a} \cdot \G(\mathbf{r},\mathbf{r}',\omega)\cdot \mathbf{b} = \mathbf{b} \cdot \G(\mathbf{r}',\mathbf{r},\omega)\cdot \mathbf{a}
\end{equation}
for arbitrary vectors $\mathbf{a}$ and $\mathbf{b}$, as can easily be proved via index notation. In fact, the above relation can be generalised from a single Green's tensor depending on two positions $\mathbf{r}$ and $\mathbf{r}'$ to $N/2$ Green's tensors depending on $N$ positions:
\begin{align}\label{reciprocityWithVectorsGeneralised}
    &\mathbf{a} \cdot \G(\mathbf{r}_1,\mathbf{r}_2,\omega)\cdot \ldots \cdot \G(\mathbf{r}_{N-1},\mathbf{r}_N,\omega)\cdot  \mathbf{b} = \notag \\
    &\quad \mathbf{b} \cdot \G(\mathbf{r}_N,\mathbf{r}_{N-1},\omega)\cdot \ldots \cdot \G(\mathbf{r}_2,\mathbf{r}_1,\omega)\cdot  \mathbf{a}
\end{align}
which we shall use later when taking the reciprocal limits of nonreciprocal quantities.  Multiplying out the modulus-square we have;
\begin{align}\label{gammaExpandedNR}
    \Gamma=\frac{2\pi \mu_0^2 \omega_\mathrm{D}^4}{\hbar}&(\dA^\downarrow \cdot \G(\rA,\rD,\omega_\mathrm{D}) \cdot \dD^\uparrow) \notag \\
    \cdot &(\dA^\uparrow \cdot \G^*(\rA,\rD,\omega_\mathrm{D}) \cdot \dD^\downarrow).
\end{align}
Using the reciprocity property \eqref{reciprocityWithVectors}, we can directly compare this to the rate of interaction $\Gamma^\mathrm{R}$ for reciprocal media;
\begin{align}\label{gammaExpandedR}
    \Gamma^\mathrm{R}=\frac{2\pi \mu_0^2 \omega_\mathrm{D}^4}{\hbar}&(\dD^\downarrow \cdot \G(\rD,\rA,\omega_\mathrm{D}) \cdot \dA^\uparrow)  \notag \\
    \cdot &(\dA^\uparrow \cdot \G^*(\rA,\rD,\omega_\mathrm{D}) \cdot \dD^\downarrow),
\end{align}
with the difference lying in the first bracketed term. This rate has a neat physical interpretation when read from right to left: the donor dipole relaxes ($\dD^\downarrow$), transmits its energy to the acceptor dipole [$\G^*(\rA,\rD,\omega_\mathrm{D})$], which excites ($\dA^\uparrow$), and then the reverse process happens. Such an interpretation cannot be made for the rate \eqref{gammaExpandedNR} in nonreciprocal media. In other words, while the form of the matrix element \eqref{eq:MfiNRSimp} remains unchanged in terms of $\G$, its modulus square appearing in Fermi's Golden Rule (and therefore the rate) cannot be simplified and interpreted in the same way as for reciprocal media.


\section{Test and validation}\label{weakNonRecipSection}

\begin{figure}[h!]
\includegraphics[width = \columnwidth]{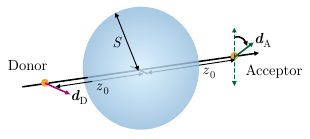}
\caption{Illustration of the geometry used for test and verification}\label{testGeomIllustration}
\end{figure}

In order to verify our formulae and assumptions, as well as the finite-difference code we will use later on for inverse design of nonreciprocal media, we first conduct a simple calculation where donor and acceptor are placed either side of a finite \new{sphere} of a nonreciprocal medium as shown in Fig.~\ref{testGeomIllustration}.
This situation cannot \new{feasibly} be described exactly analytically, so we have developed an approach to this based on \emph{weak} non-reciprocity and the Born series for the Green's tensor (see, e.g., \cite{buhmannBornExpansionCasimirPolder2006}). This approach is outlined in Appendix \ref{bornSeriesAppendix}. The resulting approximate $\G$ is:
\begin{align}\label{approximateG}
\G(\mathbf{r},&\mathbf{r}',\omega) \approx  \,\G_\mathrm{R}(\mathbf{r},\mathbf{r}',\omega)\notag \\
&+ \frac{\omega^2}{c^2}\int_V d^3 \mathbf{s} \G_\mathrm{R}(\mathbf{r},\mathbf{s}, \omega) \cdot  \delta\eps(\omega) \cdot \G_\mathrm{R}(\mathbf{s},\mathbf{r}', \omega),
\end{align}
where $\G_\mathrm{R}$ is the Green's tensor of a known, reciprocal, background permittivity distribution, and $\delta\eps$ is the tensor-valued difference between the background permittivity $\eps_\mathrm{R}$ and the true (nonreciprocal) permittivity $\eps$ of the object at hand; $\delta\eps = \eps - \eps_\mathrm{R}$. For our test case we will use the the simplest possible background permittivity, namely vacuum; $\eps_\mathrm{R} = \mathbb{I}$ --- we denote the corresponding Green's tensor as $\G^{(0)}$ and give its full form in Appendix \ref{vacGreensAppendix}. The nonreciprocal constant permittivity we use for our \new{sphere} is;
\begin{equation}\label{epsilonMatrix}
 \eps =    \begin{pmatrix}
\varepsilon_{xx} & \i \varepsilon_{xy} & 0\\
-\i \varepsilon_{xy} & \varepsilon_{xx} & 0\\
0 & 0 & \varepsilon_{zz}
\end{pmatrix}
\end{equation}
which arises from a magnetic field $\mathbf{B}_0$ aligned in the $z$ direction. Substituting the permittivity tensor \eqref{epsilonMatrix} and the vacuum Green's tensor \eqref{eq:bulk_tenserino} into Eq.~\eqref{approximateG} and carrying out the integral over the \new{sphere}, and then in turn substituting that result into Eq.~\eqref{rateNonRecipAnalytic} yields the rate of energy transfer from donor to acceptor.

We will compare the results of the above approach with finite-difference time-domain (FDTD) simulations of the corresponding system using the open-source FDTD solver MEEP~\cite{Meep}. We exploit its built-in gyrotropic Drude-Lorentz Model to define our nonreciprocal material, in which the $\eps$ tensor components are given by;
\begin{align}
    \varepsilon_{xx} &= \varepsilon_{\infty} + \frac{\omega_n^2 \Delta_n}{\Delta_n^2 - \omega^2 b^2} \sigma_n \\
     \varepsilon_{xy} &=\frac{\omega_n^2 \omega b }{\Delta_n^2 - \omega^2 b^2} \sigma_n\\
     \varepsilon_{zz} &= \frac{\omega_n^2 \sigma_n}{\Delta_n} 
\end{align}
where $\Delta^2 = \omega_n^2 - \omega^2 + \i \omega \gamma_n$,  $\varepsilon_{\infty}$ is the background permittivity of the nonreciprocal medium (not to be confused with the `background' medium the nonreciprocal object sits in), $\omega_n$ is a resonance frequency, $\gamma_n$ is a damping rate and $\sigma_n$ controls the degree of non-reciprocity. In order to calculate $\G$ using FDTD, we note that the $ij$ components of a Green's tensor $\G(\mathbf{r},\mathbf{r}',\omega)$ are deduced from the $i$th component of an electric field at $\mathbf{r}$ stemming from the $j$th component of a point current source at $\mathbf{r}'$ as (see, e.g., ~\cite{QEDBennett_2020}):
\begin{equation}\label{Gsource}
 G_{ij}(\mathbf{r},\mathbf{r}',\omega)=\frac{E_i(\mathbf{r},\omega)}{i\mu_0\omega j_j(\mathbf{r}',\omega)}, 
\end{equation}
where $\mathbf{j}(\mathbf{r},\omega)$ is the source current in the frequency domain. We implement the current as a short Gaussian pulse of central wavelength in the telecom region $\lambda_0=1.55\mu$m, the results are Fourier-transformed to the frequency domain according to the procedure detailed in~\cite{QEDBennett_2020}.

\begin{figure}
\includegraphics[width = \columnwidth]{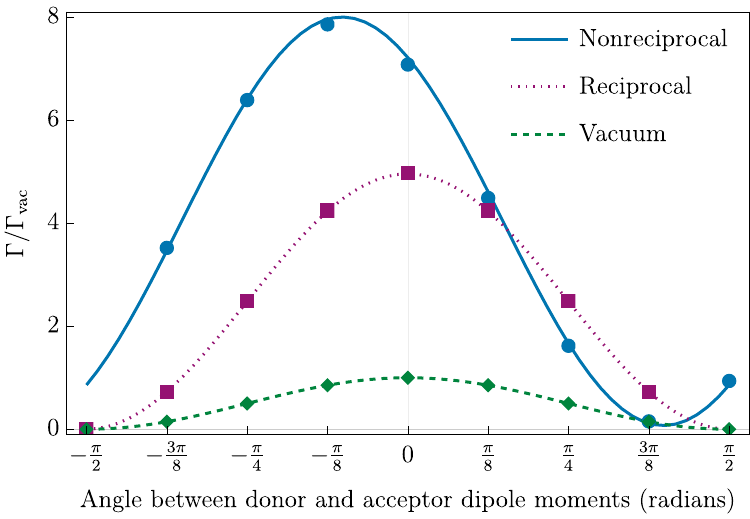}
\caption{\label{fig:gyrocurve} Comparison of RET rate using FDTD (points) and using the Born series approach \new{(lines)}. The parameters for the nonreciprocal medium are $\varepsilon_\infty = 1.444$, $\omega_n = 2.32 \omega$ (with $\omega$ being the angular frequency corresponding to the donor transition wavelength $\lambda_0 = 1.55\mu$m), \new{$b =1.855$}, $\sigma_m = 0.1$, $\gamma_n = 10^{-6}$ in the system of natural units defined by MEEP~\cite{Meep} with a length scale of one micron (although we emphasise that the scale-invariance of Maxwell's equations make this choice somewhat arbitrary). This results in off-diagonal elements of $\bar{\eps}$ being given by $\varepsilon_{xy}  \approx 0.2$ at the donor frequency. For the reciprocal medium, $\varepsilon=1.3$. The donor and acceptor are placed on the $z$ axis at $z=\pm z_0 = \pm 1.5 \mu$m, either side of a \new{sphere} of radius $S=1\mu$m (see Fig.~\ref{testGeomIllustration}). The wavelength of the donor transition is $\lambda_0 = 1.55\mu$m. All three results are normalised to the case for parallel donor and acceptor dipole moments in vacuum.}
\end{figure}

We fix the donor dipole moment to be in the $x$ direction, and allow the acceptor dipole moment to rotate in the $x-y$ plane as illustrated in Fig.~\ref{testGeomIllustration}. We normalise our all rates of transfer to that between parallel donor and acceptor in vacuum, which renders the final results independent of the magnitude of the dipole moment vectors. An example result comparing the FDTD and Born series approaches is shown in Fig.~\ref{fig:gyrocurve}, \new{where the signature of non-reciprocity is that the peak value of the rate is no longer found when donor and acceptor are parallel (somewhat analogous to a polarization rotation in a magneto-optical isolator). Our numerical and analytical results show close agreement. The slight difference between the numerics and analytics for the nonreciprocal case likely comes from the fact that our analytical approach is perturbative, only working for sufficiently weak non-reciprocity. }


\section{Inverse design using Green's tensors}\label{inverseDesignGExplanationSection}

Equation \eqref{rateNonRecipAnalytic} shows us that dependence of the rate of energy transfer upon geometry and material response can be entirely encapsulated within the Green's tensor $\G$, and that this conclusion holds in the presence of nonreciprocal media. It therefore follows that \emph{design} of the rate with nonreciprocal media effectively reduces to choosing the correct $\G$ such that the right hand side of Eq.~\eqref{rateNonRecipAnalytic} is maximised. This is exactly the approach taken in \cite{QEDBennett_2020}, in which the introduced formulae were applied to resonance energy transfer with reciprocal media. The equations introduced in \cite{QEDBennett_2020} are very general, and have been applied to different observables (e.g. environment-induced coherence \cite{bennettInverseDesignEnvironmentinduced2021}, response of metasurfaces \cite{capersDesigningCollectiveNonlocal2021} and cloaks for entanglement generation  \cite{miguel-torcalInversedesignedDielectricCloaks2022b}). 

The underlying assumption in the formalism introduced in \cite{QEDBennett_2020} was that the media involved are all reciprocal, i.e. that $\G(\mathbf{r},\mathbf{r}',\omega) = \G^\mathrm{T}(\mathbf{r}',\mathbf{r},\omega)$. Here we need to relax that assumption, so we will briefly recapitulate the contents of \cite{QEDBennett_2020} to illustrate exactly how and why the methods diverge. The goal of the optimization process is to increase the value $F$ of some functional $f$ of the Green's tensor $\G(\mathbf{r},\mathbf{r}',\omega)$
\begin{equation}\label{basicFunctional}
    F = f[\G(\mathbf{r},\mathbf{r}',\omega)].
\end{equation}
As shown in  \cite{QEDBennett_2020}, the variation $\delta F$ of this with respect to a perturbation of the environment that causes a change $\delta \G$ in the Green's tensor can  be written as;
\begin{equation}\label{perturbationInitialStep}
    \delta F = 2 \mathrm{Re}\left[ \frac{\partial f}{\partial \G}(\mathbf{r},\mathbf{r}',\omega) \odot \delta \G(\mathbf{r},\mathbf{r}',\omega) \right],
\end{equation}
where $\odot$ is the Frobenius product \footnote{For matrices $A$ and $B$,  $A \odot B \equiv \sum_{ij} A_{ij}B_{ij}$}. If the functional $f$ were to depend on multiple Green's tensors, then the variation \eqref{perturbationInitialStep} would simply be the sum of the variations with respect to the individual Green's tensors. The change $\delta \G(\mathbf{r},\mathbf{r}',\omega)$ can be rewritten via a truncated Born series;
\begin{align}\label{BornSeries}
\delta \G(\mathbf{r},\mathbf{r}',\omega )= \mu_0 \omega^2 & \int_V d^3  \mathbf{r}''  n(\mathbf{r}'')\alpha(\mathbf{r}'')\notag \\
&\times \G(\mathbf{r},\mathbf{r}'',\omega ) \cdot \G(\mathbf{r}'',\mathbf{r}',\omega),
\end{align}
where the volume $V$ runs over the (small) region of the environment that has changed, $n(\mathbf{r})$ is the number density of atoms within that volume and $\alpha(\mathbf{r})$ are their polarisabilities.  Assuming that the number density and polarisability do not vary over the perturbation region and using Eq.~\eqref{BornSeries} in Eq.~\eqref{perturbationInitialStep}, we have;
\begin{align}
    \delta F =& 2 \alpha n \mathrm{Re}\int_V d^3  \mathbf{r}'' \notag \\
    &\times \frac{\partial f}{\partial \G}(\mathbf{r},\mathbf{r}',\omega) \odot \big[  \G(\mathbf{r},\mathbf{r}'',\omega ) \cdot \G(\mathbf{r}'',\mathbf{r}',\omega) \big].
\end{align}
We can drop the (positive) constants $2\alpha n$ since all that will turn out to matter is the maximum of this as a function of the choice of perturbation, so we are left with;

\begin{align}\label{perturbationWithBornSeries}
    \delta F =& \mathrm{Re}\int_V d^3  \mathbf{r}''  \frac{\partial f}{\partial \G}(\mathbf{r},\mathbf{r}',\omega) \odot \big[  \G(\mathbf{r},\mathbf{r}'',\omega ) \cdot \G(\mathbf{r}'',\mathbf{r}',\omega) \big].
\end{align}
\new{In order to implement the optimization via Eq.~\eqref{perturbationWithBornSeries}} we will employ a powerful and general method called the level-set approach \cite{OSHER198812}. In the level-set method the medium is described by a level-set function $\Phi$ whose zero-level contour $\Phi(t)=0$ corresponds to the boundary of the medium, which in this case will be nonreciprocal. The initial shape of the boundary is defined by 
\begin{equation}\label{levelSetDef}
    \Phi(\mathbf{r}(t), t) = 0,
\end{equation}
where an artificial `time' parameter $t$ has been introduced, representing the iterative update process, and $\Phi<0$ inside the medium ($\Phi>0$ outside). The total time derivative of \eqref{levelSetDef} leads to an advection equation governing the evolution of the boundary: 
\begin{equation}\label{Eq:adv}
 \frac{\partial \Phi}{\partial t}+\mathbf{v}\cdot\nabla\Phi=0 .
\end{equation}
Considering the normal vector to the boundary $\mathbf{n}=\nabla\Phi/\left|\nabla\Phi\right|$, Eq.~\eqref{Eq:adv} can be re-written as:
\begin{equation}\label{Eq:levelset}
 \frac{\partial \Phi}{\partial t}+v_n\left|\nabla\Phi\right|=0,  
\end{equation}
where $v_n=\mathbf{v}\cdot\mathbf{n}$ is the scalar velocity field in the direction normal direction of the boundary. We need to choose the velocity field $v_n$ such that the boundary deforms in a way that $\delta F$ is positive (and therefore $F$ increases). As discussed in \cite{deformationpaper}, the correct choice of velocity field can be found by rewriting the integration over $\mathbf{r}''$ in Eq.~\eqref{perturbationWithBornSeries} as;
\begin{equation}\label{levelSetIntegralsAlgebra}
\int_V d^3  \mathbf{r}'' \to \int_{\partial V} d A \delta x(\mathbf{r}'') = \int_{\partial V} d A v_n \delta t,
\end{equation}
where $\delta x$ is the size of an infinitesimal deformation perpendicular to the boundary, and the integral is now over its area $A$. In the final step of Eq.~\eqref{levelSetIntegralsAlgebra} we have replaced the perpendicular deformation with the product of an infinitesimal time step and the velocity perpendicular to the boundary. Using this in Eq.~\eqref{perturbationWithBornSeries} we have;
\begin{align}
    &\delta F =  \mathrm{Re}\int_{\partial V} d A v_n \delta t \notag \\
    &\times \frac{\partial f}{\partial \G}(\mathbf{r},\mathbf{r}',\omega) \odot \big[  \G(\mathbf{r},\mathbf{r}'',\omega ) \cdot \G(\mathbf{r}'',\mathbf{r}',\omega) \big].
\end{align}

Analogously to the reciprocal case discussed in \cite{QEDBennett_2020} and \cite{deformationpaper}, we can guarantee positive change in the merit function by choosing the velocity field $v_n$ to be;
\begin{equation}\label{vnnumerical}
v_n=\mathrm{Re}\left\{\frac{\partial f}{\partial \G}(\mathbf{r},\mathbf{r}',\omega) \odot \big[  \G(\mathbf{r},\mathbf{r}'',\omega ) \cdot \G(\mathbf{r}'',\mathbf{r}',\omega) \big]\right\},
\end{equation}
since this means that $\delta F =\int_{\partial V} d A v_n^2 $, which is positive. The particular form of the above function depends on the choice of observable (and thereby choice of functional $f$), which we will discuss in the next section. 

In principle, Eq.~\eqref{vnnumerical} is enough to begin an optimization. In practice, the problem is that the (variable) optimization position $\mathbf{r}''$ appears in the \emph{second} argument in one of the Green's tensors. The second argument of a Green's tensor corresponds to the source, so an optimization must consider  each `candidate' position for the perturbation via a separate simulation (i.e.~with different sources). There may be overwhelmingly many of these in a large-scale 3D problem, so a trick is required in order to make the scheme numerically feasible. 

In reciprocal media, the problem is solved simply by taking advantage of reciprocity to write $\G(\mathbf{r},\mathbf{r}'',\omega ) = \G^\mathrm{T}(\mathbf{r}'',\mathbf{r},\omega )$, giving;
\begin{align}
    \delta & F_\mathrm{recip} = \mathrm{Re}\int_V d^3  \mathbf{r}'' \notag \\
    &\times \frac{\partial f}{\partial \G}(\mathbf{r},\mathbf{r}',\omega) \odot \big[  \G^\mathrm{T}(\mathbf{r}'',\mathbf{r},\omega ) \cdot \G(\mathbf{r}'',\mathbf{r}',\omega) \big],
\end{align}
in which case all the positions appearing in the `source' arguments in all the Green's tensors are the fixed, physical source or observation points $\mathbf{r}$ or $\mathbf{r}'$. This brings the number of required simulations down to \emph{two}, regardless of the number of candidate optimization positions $\mathbf{r}''$. This is essentially the well-known adjoint method, but expressed in a particularly elegant and direct way \cite{QEDBennett_2020}. 

In nonreciprocal media this route to solving the problems with  relation \eqref{perturbationWithBornSeries} is not open to us since, in nonreciprocal media $\G(\mathbf{r},\mathbf{r}',\omega) \neq \G^\mathrm{T}(\mathbf{r}',\mathbf{r},\omega)$. We can, however, use a separate but related property of the Green's tensor of a Faraday medium (a particular case of a nonreciprocal medium), namely that \cite{casimirOnsagerPrincipleMicroscopic1945,villeneuveReciprocityRelationshipsGyrotropic1958}; 
\begin{equation}\label{FaradayAdjoint}
    \G(\mathbf{r},\mathbf{r}',\omega; \mathbf{B}_0) = \G^\mathrm{T}(\mathbf{r},\mathbf{r}',\omega; -\mathbf{B}_0),
\end{equation}
where the additional argument $\mathbf{B}_0$ represents the external applied field. Thus for Faraday media we can introduce a slightly modified version of the adjoint method based on the symmetry implied by Eq.~\eqref{FaradayAdjoint}, rather than the symmetry of reciprocity. Using Eq.~\eqref{FaradayAdjoint} in Eq.~\eqref{perturbationWithBornSeries} gives;
\begin{align}\label{dFFaraday}
    \delta & F_\mathrm{Faraday} =  \mathrm{Re}\int_V d^3  \mathbf{r}''\frac{\partial f}{\partial \G}(\mathbf{r},\mathbf{r}',\omega;\mathbf{B}_0) \notag \\
    & \qquad  \odot \big[  \G^\mathrm{T}(\mathbf{r}'',\mathbf{r},\omega ;-\mathbf{B}_0) \cdot \G(\mathbf{r}'',\mathbf{r}',\omega;\mathbf{B}_0) \big],
\end{align}
with the corresponding boundary velocity being
\begin{align}\label{vnnumericalFaraday}
v_n=\mathrm{Re}\Bigg\{\frac{\partial f}{\partial \G}(\mathbf{r},\mathbf{r}',\omega;\mathbf{B}_0) &\odot \big[  \G^\mathrm{T}(\mathbf{r}'',\mathbf{r},\omega ;-\mathbf{B}_0)  \notag \\
&\cdot \G(\mathbf{r}'',\mathbf{r}',\omega;\mathbf{B}_0) \big]\Bigg\}.
\end{align}
This expression has the required quality of the optimization position $\mathbf{r}''$ appearing only in the first argument of Green's tensors, so can be regarded as an analog of the adjoint method but applicable to non-recipriocal media. Since this is not quite the adjoint method, we will refer to it as the `Faraday-adjoint' method in the remainder of this work.


\section{Application: inverse design of RET isolation}\label{demonstrationSection}

As discussed in the introduction, magneto-optical isolators are the photonic analogues of electrical diodes, they enable unidirectional propagation of light.  They are used to protect laser sources from back reflections detrimental to their performance, and are often based on magneto-optical media. In this section, we present what is, to the best of our knowledge, the first instance of inverse designed magneto-optical (nonreciprocal) media, in which we will produce one-way RET from donor to acceptor as schematically illustrated in Fig.~\ref{generalRETFig}. 

Based on Eq.~\eqref{rateNonRecipAnalytic}, the RET-isolation strength of a magneto-optical isolator can be expressed by the means of a merit function $R$: 
\begin{equation}\label{ratio}
\merit=\frac{\Gamma_+}{\Gamma_-}=\frac{\left|\dA\cdot \G(\rA,\rD)\cdot \dD\right|^2}{\left|\dD\cdot \G(\rD,\rA)\cdot \dA\right|^2}    
\end{equation}
where $\Gamma_+$ and $\Gamma_-$ are the resonance energy transfer rates for forward and backward transfer, respectively, between a donor dipole D and an acceptor dipole A. We have dropped the frequency argument for brevity --- from now on all Green's tensors should be assumed to be evaluated at the donor frequency $\omega_\mathrm{D}$.  Applying the reciprocity relation \eqref{reciprocityWithVectors} to the isolation ratio \eqref{ratio} of course produces $\merit = 1$ --- reciprocal (linear, passive) media cannot provide isolation.

We emphasise that the Green's tensors $\G(\mathbf{r},\mathbf{r}')$ and $\G(\mathbf{r}',\mathbf{r})$ are, in principle, \emph{unrelated} when a nonreciprocal medium (of any type, not necessarily a Faraday medium) is present. Thus, the very general functional we begin with is
\begin{equation}
    F = f[\G(\mathbf{r},\mathbf{r}'),\G(\mathbf{r}',\mathbf{r})],
\end{equation}
which, as noted below Eq.~\eqref{perturbationInitialStep}, means that the variation becomes the sum of the variations with respect to the individual Green's tensors. Recapitulating the derivation from Eq.~\eqref{basicFunctional} to \eqref{vnnumerical} under these conditions, the boundary velocity of the medium of the merit function with $\G$ then reads: 
\begin{align}\label{basic}
v_n&=\text{Re}\bigg\{\frac{\partial \merit}{\partial\G}(\rA,\rD)\odot [\G(\rA,\rOpt)\cdot\G(\rOpt,\rD)]\nonumber\\&+\frac{\partial\merit}{\partial\G}(\rD,\rA)\odot [\G(\rD,\rOpt)\cdot\G(\rOpt,\rA)]\bigg\}.
\end{align}
The main algebraic task in readying the above equation for numerical evaluation is calculation of the functional derivatives of the ratio $\merit$ shown in Eq.~\eqref{ratio} with respect to $\G(\rA,\rD)$ and $\G(\rD,\rA)$. This is tedious but straightforward \new{(see Appendix \ref{derivationAppendix})}, producing:
\begin{widetext}
\begin{align}\label{eq:dFNoFlip}
v_n=&\text{Re}\bigg\{\frac{(\dA\cdot \G^{*}(\rA,\rD)\cdot\dD)[\dA\cdot \G(\rA,\rOpt)\cdot \G(\rOpt,\rD)\cdot \dD] }{\left| \dD\cdot \G(\rD,\rA)\cdot \dA\right|^2}\nonumber\\
&\qquad \qquad \qquad\qquad\qquad -\frac{\left|\dA\cdot \G(\rA,\rD)\cdot \dD\right|^2(\dD\cdot \G(\rD,\rOpt)\cdot \G(\rOpt,\rA) \cdot \dA)}{[\dD\cdot \G^*(\rD,\rA)\cdot \dA]\left|\dD\cdot \G(\rD,\rA)\cdot \dA\right|^2}\bigg\}\;.
\end{align}
Applying the reciprocity relation $\G(\rA,\rD) = \G^\mathrm{T}(\rA,\rD)$ via Eq.~\eqref{reciprocityWithVectorsGeneralised} to the above produces $v_n = 0$ as it must, since it is the derivative of Eq.~\eqref{ratio} which is constant in the same limit.  As outlined in Section \ref{inverseDesignGExplanationSection}, calculation of $\G(\mathbf{r}_{\mathrm{A/D}},\rOpt)$ is computationally expensive. However, this problem can be sidestepped by exploiting the Faraday-adjoint relation \eqref{FaradayAdjoint}, yielding;
\begin{align}\label{eq:dF}
v_n=&\text{Re}\bigg\{\frac{(\dA\cdot \G^{*}(\rA,\rD;\mathbf{B}_0)\cdot\dD)[\dA\cdot \G^\mathrm{T}(\rOpt,\rA;-\mathbf{B}_0)\cdot \G(\rOpt,\rD;\mathbf{B}_0)\cdot \dD] }{\left| \dD\cdot \G(\rD,\rA;\mathbf{B}_0)\cdot \dA\right|^2}\nonumber\\
&\qquad \qquad \qquad\qquad\qquad -\frac{\left|\dA\cdot \G(\rA,\rD;\mathbf{B}_0)\cdot \dD\right|^2[\dD\cdot \G^\mathrm{T}(\rOpt,\rD;-\mathbf{B}_0)\cdot \G(\rOpt,\rA;\mathbf{B}_0) \cdot \dA]}{[\dD\cdot \G^*(\rD,\rA;\mathbf{B}_0)\cdot \dA]\left|\dD\cdot \G(\rD,\rA;\mathbf{B}_0)\cdot \dA\right|^2}\bigg\}\;.
\end{align}
\end{widetext}
Now, all optimization positions $\rOpt$ appear in the first (i.e., observation) argument of all Green's tensors which, as discussed in Section  \ref{inverseDesignGExplanationSection} and extensively in Ref.~\cite{QEDBennett_2020}, allows for an efficient optimization process. \new{We note that the introduction of a bias vector $\mathbf{B}_0$ is only one possibility from a broader class of any bias vector that changes under time reversal. One example of a different type of bias vector would be the angular momentum vector arising from using a rotating medium --- this would be an example of the mechanical Faraday effect \cite{nienhuisMagneticMechanicalFaraday1992}.}

The right hand side of Eq.~\eqref{eq:dF} can be determined for any geometry using the FDTD procedure outlined in Section \ref{weakNonRecipSection} by calculating four Green's tensors 
$\G(\rOpt,\rD;\mathbf{B}_{0})$, $\G(\rOpt,\rA;\mathbf{B}_{0})$, $\G(\rOpt,\rD;-\mathbf{B}_{0})$ and $\G(\rOpt,\rA;-\mathbf{B}_{0})$ --- the Faraday-adjoint method is twice as computationally intensive as the adjoint method where only two Green's tensors are required. In the case of the latter two Green's tensors, the direction of the bias vector is reversed to change the  properties of the Faraday medium while maintaining its geometry, which is a physically distinct situation to that with the original field orientation, in effect `cancelling out' the distinction introduced by swapping the positions of donor and acceptor.

We will use Eq.~\eqref{eq:dF} to perform the inverse design of magneto-optical RET isolation, since increasing $R$ as defined by Eq.~\eqref{ratio} corresponds \new{to an increased RET isolation strength}. We use the same \new{sphere} considered in the test simulations shown in Fig.~\ref{fig:gyrocurve} as our starting geometry, with a vacuum background. 

In order to align this calculation as much as possible with the ideas required to develop a technological isolator, we ensure that our donor and acceptor are sufficiently far apart that the maximal rate of energy transfer is found when their dipole moments are parallel to each other and perpendicular to their separation vector (and minimised when they are perpendicular and perpendicular to their separation vector). This is based on the idea of filtering out an orthogonal reflected polarization in a realistic isolator, and is closely related to being in the far-field regime --- the ideas of polarization and travelling waves are not well-defined in the near-field regime. In fact, if the donor and acceptor are too close, the maximal rate is found when the dipoles are placed end-to-end (i.e. parallel to each other \emph{and} to their separation vector), with the crossover point appearing at a separation distances of $\frac{1}{2 \pi }\sqrt{\frac{1}{2} \left(\sqrt{37}+5\right)}\approx 0.37$ times the wavelength of the donor transition $\lambda_0$ (see, e.g., Fig 1d of \cite{rustomjiDirectImagingEnergyTransfer2019} for the same quantity expressed in terms of wavenumber). Since our donor and acceptor are $3\mu$m apart and the transition wavelength is $1.55\mu$m the separation is $\approx 1.9$ times the wavelength, meaning we are well within the desired regime. The dipole moments in our optimization are $\dD=|\dD|(1,0,0)$ and $\dA=|\dA|(\frac{1}{\sqrt{2}},\frac{1}{\sqrt{2}},0)$, this amounts to a $45^\circ$ rotation signature typical of conventional Faraday isolators~\cite{Gauthier:86} based on polarization rotation of a input wave.

\begin{figure}[h!]
\includegraphics[width = 0.45\textwidth]{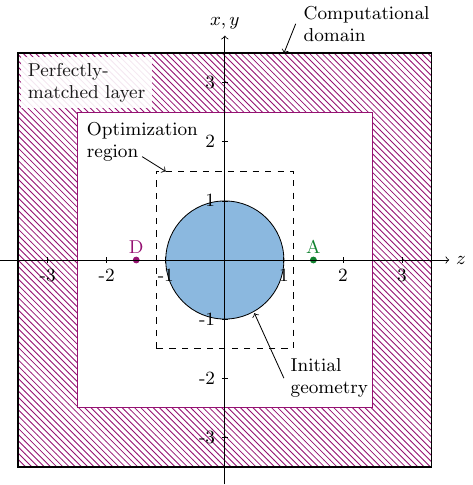}
\caption{Computational setup. The computational domain is bounded by perfectly matched layers, whose non-reflecting property ensures no spurious reflections from the boundaries. Within that we define a region between donor and acceptor into which the algorithm is allowed to deform the nonreciprocal medium. We choose the initial geometry to be the same as used in the test case discussed in Section \ref{weakNonRecipSection} and shown in Fig.~\ref{testGeomIllustration}. The distance unit is $\mu \text{m}$. 
}\label{computationalSetup}
\end{figure}

The computational domain is a cube of side length $7\mu\mathrm{m}$, as illustrated in Fig.~\ref{computationalSetup}. Perfectly matched layers \cite{berengerPerfectlyMatchedLayer1994} of thickness $1\mu\mathrm{m}$ are placed at the borders, leaving a usable volume of $(5\mu\mathrm{m})^3$. The design volume in which the Faraday material can be engineered is chosen as an origin-centered cuboid of side length $3\mu\mathrm{m}$ in the $x$ and $y$ directions, and  $(4/3)\mu \mathrm{m}$ in the $z$ to ensure that a minimal distance of $(2/3)\mu\mathrm{m}>0.37\lambda_0$ always separates the medium from the dipoles.

While we rely on MEEP to carry out finite difference time domain simulations and to simulate the Faraday medium, we perform topology optimization using a custom-made 3D algorithm.   
As discussed in Section \ref{inverseDesignGExplanationSection}, the structural domain is represented by a level-set function $\Phi$ whose zero-level contour $\Phi(t)=0$ corresponds to the boundary of the Faraday medium. This iterative process is implemented as follows: given an initial geometry, we calculate the velocity on the boundary as given by Eq.~\eqref{eq:dF}. The velocity must then be formally extended to the rest of the domain~\cite{sethian1999level}, to do this use we use a Python module \texttt{scikit-fmm}~\cite{scikit} to compute the signed distance function from the zero level set contour. 

Once the velocity field has been obtained throughout the whole domain, the boundary can be updated using the advection equation \eqref{Eq:levelset}. We solve Eq.~\eqref{Eq:levelset} using the upwind scheme from a Python PDE solver module \texttt{FIPY} \cite{fipy}. We take advantage of the signed distance function to limit the maximum advection distance to a few pixels in order to prevent the boundary from deforming excessively [i.e. beyond the limits of applicability of the truncated Born series \eqref{BornSeries}]. 
\new{The Courant–Friedrichs–Lewy stability condition for the advection solver $C$~\cite{hirsch2007numerical} evaluates to less than unity: we used $C=0.7$.}

\begin{figure}
\includegraphics{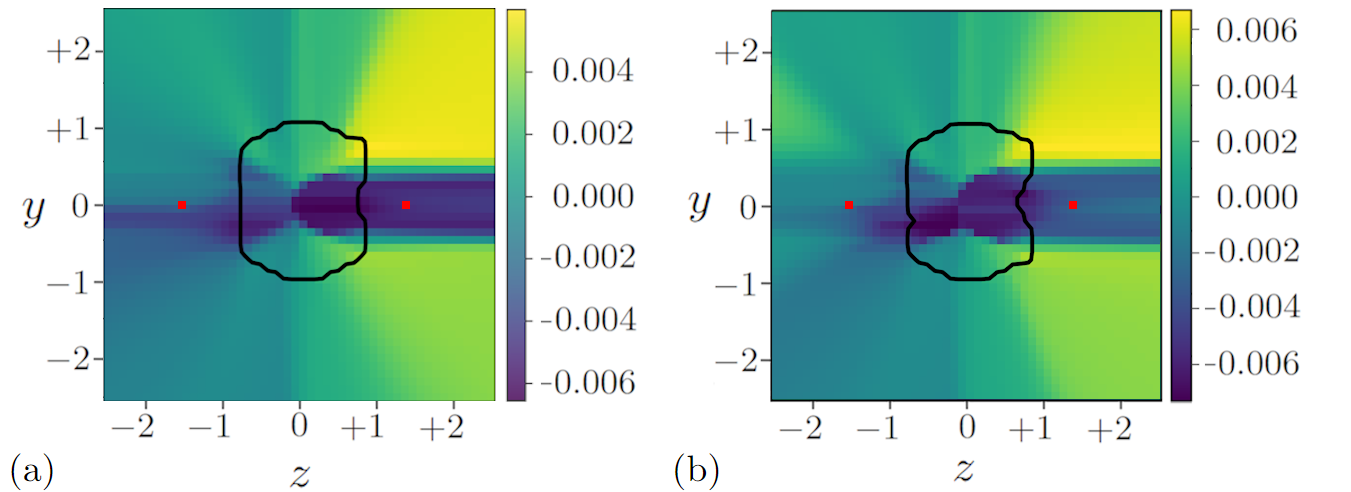}
\caption{\new{(a) Iteration 3 and (b) iteration 4 of the advected geometry}. The boundary (zero-level set) of the Faraday medium is represented by a solid black line. The red dots indicate the position of the acceptor and donor dipoles. The extended velocity field $v_{ext}$ is shown as a coloured background. \secondNew{The distance unit is $\mu \text{m}$}.}\label{figvel}
\end{figure}

\begin{figure}
\includegraphics{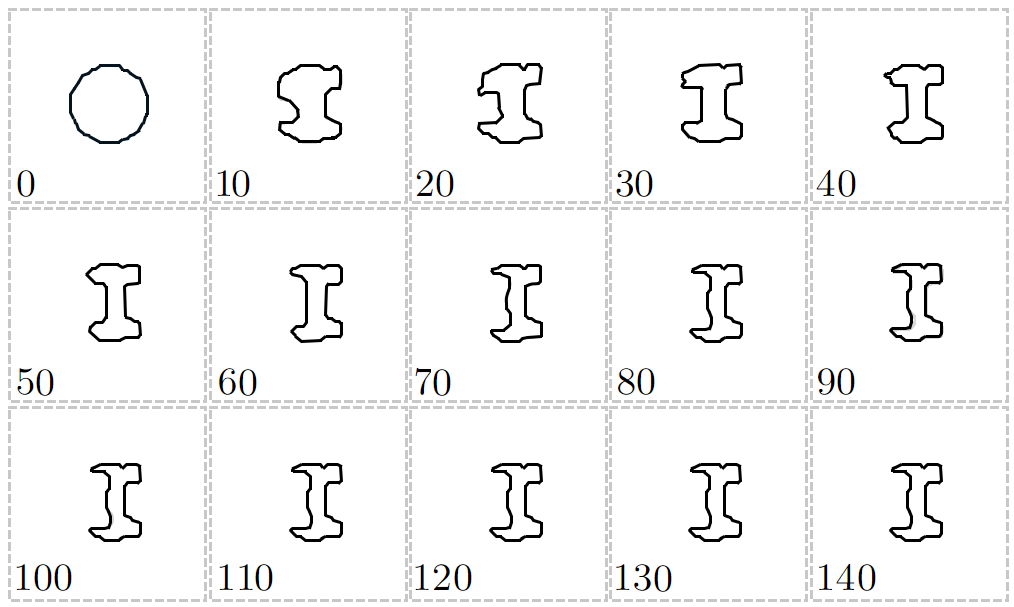}
\caption{Slices of the advected geometry at $x=0$ showing the evolution of the boundary (zero-contour) as the iteration number increases.}\label{fig:evolution}
\end{figure}

Once the boundary of the initial geometry has been advected, we obtain a new geometry which is used as the starting point for the next iteration. 
Fig.~\ref{figvel} 
shows the initial geometry, a $yz$-plane slice of the \new{sphere} of Faraday medium, and the advected geometry used as the starting point for the next iteration. The displacements of the boundary follow the pattern dictated by the extended velocity field showed in the background, reminiscent of the  way a sandcastle would be deformed by a flow of water or clouds by the flow of air. After a small amount of deformation a new velocity field is calculated and the process repeats --- the resulting evolution of the initial shape is shown in Fig.~\ref{fig:evolution}.

\begin{figure}
\includegraphics{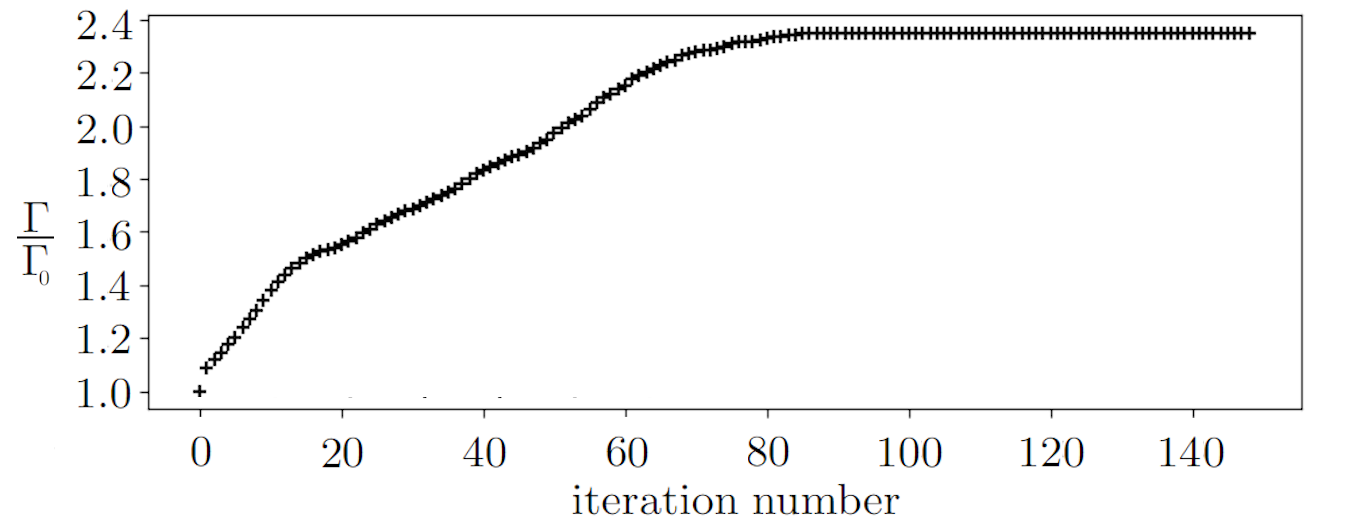}
\caption{\label{fig:final}The (normalized) isolation strength $\Gamma$ of the gyro-magnetic optical isolator increases during the iterative optimization process. $\Gamma_0$, the initial isolation strength (iteration 0) is used as the normalization values.}
\end{figure}

\begin{figure}
\includegraphics{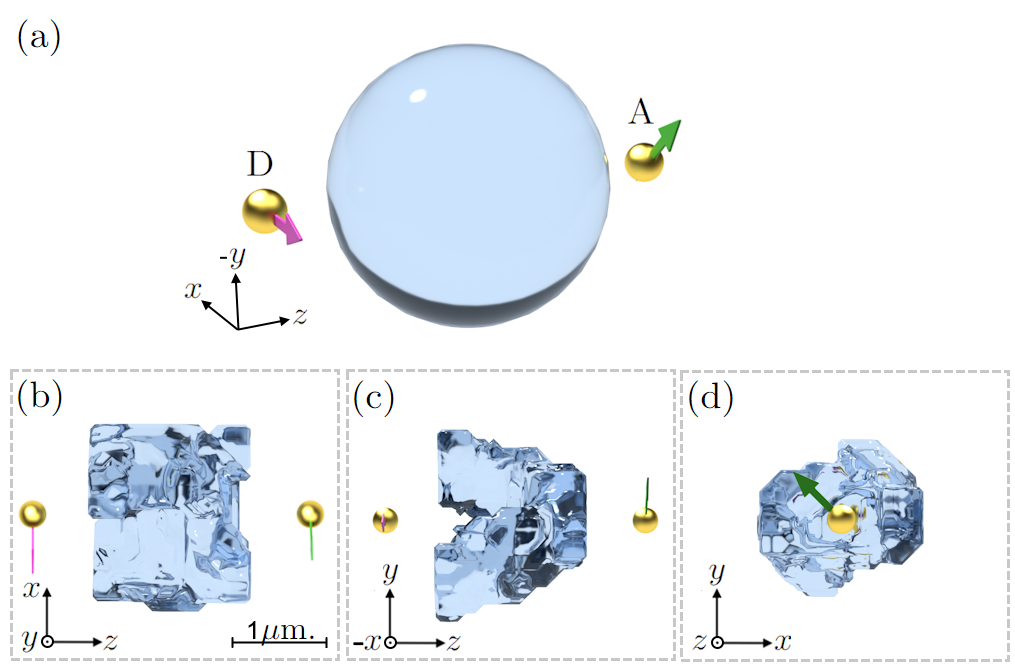}
\caption{\label{fig:geom} a) Initial geometry and b-d) final geometry of the inverse-designed nonreciprocal RET-isolator. The donor (D) and acceptor (A) dipoles are represented as gold spheres.}
\end{figure}

\new{We terminate the iteration process when the structure converges to a final shape}.  Figure \ref{fig:final}
shows the increase in the isolation strength of the magneto-optical RET-isolator as a function of the number of iterations.  Figure \ref{fig:geom}
shows the initial geometry and multiple perspectives on the final geometry of the topology-optimised Faraday medium obtained using our algorithm. In this case, the algorithm \new{begins to cut} a hole in the middle of the \new{sphere, pushing the rest of the material outwards into a ring}. Such a geometry improves the isolation strength, \new{converging to a factor of approximately 2.3 after around 80 iterations.}

Naturally, the example we have shown here serves purely illustrative purposes to demonstrate how the Green's tensor formalism can be used to perform the inverse design of isolation processes in general. \new{We caution that, in common with other inverse design processes based on the adjoint method, the final geometry produced by the optimization here does not in general represent a globally optimal design \cite{millerFundamentalLimitsExtinction2014,Molesky2018}, nor is it practical for a realistic photonic isolator}. It does, however demonstrate unequivocally that the tools we have introduced can be exploited to undertake inverse design of Faraday media, opening up a new avenue of research in the quest for an integrated photonic isolator.


\section{Conclusion}

In this work we have described and given an example implementation of a suite of tools that are required to apply the adjoint method of inverse design to nonreciprocal media. Introducing RET in non-recpiprocal media (which itself is interesting from a theoretical point of view and will be pursued in detail elsewhere), we used the example of RET-isolation to demonstrate the introduced Faraday-adjoint method, and showed that three-dimensional level-set optimization methods can produce improved devices in this context. This is a stepping stone towards the significant technological goal of an integrated optical isolator, which could take advantage of inverse design methods to use CMOS-compatible (though weakly gyrotropic) materials --- even silicon itself. Such an endeavour will build on the physical principles introduced here to incorporate engineering considerations including, for example, manufacturing constraints (its own sub-field of inverse design, see e.g. \cite{sigmundManufacturingTolerantTopology2009,vercruysseAnalyticalLevelSet2019,augensteinInverseDesignNanophotonic2020}), extended source fields and observation regions, broadband operation and CMOS integration. Even outside of this, the methods used here could find applications in improving the performance of bulk isolators based on rare-earth iron garnets.


\begin{acknowledgments}
It is a pleasure to acknowledge discussions with R.~Kilianski. C.M.C and R.B. acknowledge financial support from UK Research and Innovation Council (UKRI) grant EP/W016486/1. M.C.W acknowledges financial support from EPSRC Doctoral Training Programme grant EPSRC/DTP 2020/21/EP/T517896/1.  
\end{acknowledgments}

\appendix

\section{Born series for nonreciprocal media}\label{bornSeriesAppendix}

The generalised Helmholtz equation we show in Eq.~\eqref{generalisedHelmholtz} (and use for the remainder of that section) actually applies to media that are non-local and anisotropic (non-reciprocity being a specific case of anisotropy), with arbitrary electromagnetic susceptibilities and cross-susceptibilities. In practice for this work we only require a local, electrically anisotropic medium with no cross-susceptibilities, under which conditions Eq.~\eqref{generalisedHelmholtz} simplifies to \cite{buhmannMacroscopicQuantumElectrodynamics2012};
\begin{equation}\label{GNonRecipMedium}
\nabla \times \nabla \times \G(\mathbf{r},\mathbf{r}',\omega) - \frac{\omega^2}{c^2} \eps \cdot \G(\mathbf{r},\mathbf{r}',\omega) = \delta(\mathbf{r}-\mathbf{r}')
\end{equation}
To obtain a perturbative solution we assume that the nonreciprocal permittivity can be written as the sum of reciprocal part $\eps_\mathrm{R}$ and a nonreciprocal additional part $\delta \eps$, so that
\begin{equation}
     \eps(\mathbf{r},\omega)  = \delta \eps(\mathbf{r},\omega) + \eps_\mathrm{R}(\mathbf{r},\omega).
\end{equation}
We further assume that the Green's tensor $\G_\mathrm{R}$ for the geometry defined by the reciprocal part $\eps_\mathrm{R}$ is known, as in, we know the solution to;
\begin{equation}\label{GRecipOnly}
\nabla \times \nabla \times \G_\mathrm{R}(\mathbf{r},\mathbf{r}',\omega) - \frac{\omega^2}{c^2} \eps_\mathrm{R}(\mathbf{r},\omega) \cdot \G_\mathrm{R}(\mathbf{r},\mathbf{r}',\omega) = \delta(\mathbf{r}-\mathbf{r}').
\end{equation}
Subtracting \eqref{GRecipOnly} from \eqref{GNonRecipMedium} (and temporarily suppressing all position frequency arguments), we have;
\begin{equation}
\nabla \times \nabla \times (\G - \G_\mathrm{R})- \frac{\omega^2}{c^2} (\eps \cdot \G - \eps_\mathrm{R} \cdot \G_\mathrm{R}) = 0 .
\end{equation}
Defining
\begin{equation}\label{defintionsOfDeltaG}
\delta \G = \G - \G_\mathrm{R}  ,
\end{equation}
we aim to eliminate `full' quantities $\G$ and $\eps$ in favour of the `small' unknown quantities $\delta \G$, $\delta \eps$ and the known quantities $\G_\mathrm{R}$ and $\eps_\mathrm{R}$;
\begin{equation}
\nabla \times \nabla \times \delta \G- \frac{\omega^2}{c^2} \left[ (\delta\eps + \eps_\mathrm{R}) \cdot (\delta \G + \G_\mathrm{R}) - \eps_\mathrm{R} \cdot \G_\mathrm{R}\right] = 0 .
\end{equation}
The terms $\eps_\mathrm{R} \cdot \G_\mathrm{R}$ cancel and we are left with;
\begin{equation}
\nabla \times \nabla \times \delta \G- \frac{\omega^2}{c^2} \left[ (\delta\eps \cdot \delta \G +\delta\eps \cdot  \G_\mathrm{R} +\eps_\mathrm{R} \cdot \delta \G  \right] = 0 .
\end{equation}
Rearranging all the nonreciprocal parts onto the right hand side to act as a `source' of non-reciprocity, we have;
\begin{equation}\label{startingEqForBornSeries}
\nabla \times \nabla \times \delta \G- \frac{\omega^2}{c^2}  \eps_\mathrm{R} \cdot \delta \G   = \frac{\omega^2}{c^2} \delta\eps \cdot \left(\delta \G + \G_\mathrm{R} \right).
\end{equation}
This is an inhomogeneous vector differential equation, which can be solved by yet another Green's tensor. We will call this Green's tensor $\mathbb{H}$, defined as satisfying;
\begin{equation}\label{auxiliaryGEq}
\nabla \times \nabla \times \mathbb{H}- \frac{\omega^2}{c^2}  \eps_\mathrm{R} \cdot \mathbb{H}  = \delta(\mathbf{r}-\mathbf{r}').
\end{equation}
Comparing Eq.~\eqref{GRecipOnly} and Eq.~\eqref{auxiliaryGEq}, we see that $\G_\mathrm{R} = \mathbb{H}$. Therefore, an exact (but formal) solution to Eq.~\eqref{startingEqForBornSeries} can be written as;
\begin{align}
\delta \G(\mathbf{r},\mathbf{r}',\omega) = \int d^3 \mathbf{s} & \G(\mathbf{r},\mathbf{s}, \omega) \cdot \frac{\omega^2}{c^2} \delta\eps(\mathbf{s},\omega) \cdot \Big[\delta \G_\mathrm{R}(\mathbf{s},\mathbf{r}',\omega)\notag  \\
&+ \G_\mathrm{R}(\mathbf{s},\mathbf{r}',\omega)  \Big].
\end{align}
Or, using the definition \eqref{defintionsOfDeltaG}:
\begin{align}
\G(\mathbf{r},\mathbf{r}',&\omega) =\G_\mathrm{R}(\mathbf{r},\mathbf{r}',\omega) \notag \\
&+ \frac{\omega^2}{c^2}\int d^3 \mathbf{s} \G_\mathrm{R}(\mathbf{r},\mathbf{s}, \omega) \cdot  \delta\eps(\mathbf{s},\omega) \cdot \G(\mathbf{s},\mathbf{r}', \omega).
\end{align}
This is a self-referential equation for $\G(\mathbf{r},\mathbf{r}',\omega)$, so we solve it via repeated re-substitution --- this is the standard method of the Born series. Restricting to one non-trivial term (i.e. assuming that the nonreciprocal perturbation is small) we have;
\begin{align}
\G(\mathbf{r},\mathbf{r}',&\omega) \approx  \,\G_\mathrm{R}(\mathbf{r},\mathbf{r}',\omega)\notag \\
&+ \frac{\omega^2}{c^2}\int d^3 \mathbf{s} \G_\mathrm{R}(\mathbf{r},\mathbf{s}, \omega) \cdot  \delta\eps(\mathbf{s},\omega) \cdot \G_\mathrm{R}(\mathbf{s},\mathbf{r}', \omega),
\end{align}
which allows us to work out nonreciprocal Green's tensors with knowledge only of the Green's tensor of a corresponding `close-by' reciprocal Green's tensor, and the nonreciprocal permittivity. If we assume that our nonreciprocal object has a constant (tensor) permittivity and sits in a reciprocal background, we can take $\delta\eps(\mathbf{r},\omega)$ to have the following form;
\begin{equation}
    \delta\eps(\mathbf{r},\omega) = \begin{cases}
        \delta\eps(\omega) &\text{if } \mathbf{r} \in V \\
        0 &\text{otherwise}
    \end{cases}
\end{equation}
where $V$ is the volume of the nonreciprocal object. We can therefore restrict the $\mathbf{s}$ integration to the volume $V$, leaving:
\begin{align}
\G(\mathbf{r},\mathbf{r}',&\omega) \approx  \,\G_\mathrm{R}(\mathbf{r},\mathbf{r}',\omega)\notag \\
&+ \frac{\omega^2}{c^2}\int_V d^3 s \G_\mathrm{R}(\mathbf{r},\mathbf{s}, \omega) \cdot  \delta\eps(\omega) \cdot \G_\mathrm{R}(\mathbf{s},\mathbf{r}', \omega).
\end{align}
This is Eq.~\eqref{approximateG} in the main text. 

\section{Vacuum Green's tensor}\label{vacGreensAppendix}

The Green's tensor for infinite unbounded vacuum is (see, for example, \cite{buhmann2012Book1});
\begin{align}
\mathbb{G}^{(0)}(\mathbf{r},\mathbf{r}',\omega)=&-\frac{\mathbb{I} }{3k ^2}{\delta}^{(3)}(\boldsymbol{\rho})\notag \\
&-\frac{e^{\i k \rho}}{4\pi k ^2 \rho^3}\Bigg\{\left[1-\i k \rho-\left(k \rho\right)^2\right]\mathbb{I}\notag \\
&-\left[3-3\i k \rho -\left(k \rho\right)^2\right]\;\mathbf{e}_\rho\otimes\mathbf{e}_\rho\Bigg\},
\label{eq:bulk_tenserino}
\end{align}
where $k=\omega/c$, $\boldsymbol{\rho}=\mathbf{r}-\mathbf{r}'$, $\rho = \abs{\boldsymbol{\rho}}$, and  $\mathbf{e}_\rho=\boldsymbol{\rho}/\rho$. This is used as the `background' reciprocal Green's tensor  $\G_\mathrm{R}$ in Eq.~\eqref{approximateG} to produce the results shown in Fig.~\ref{fig:gyrocurve} via Eq.~\eqref{rateNonRecipAnalytic}. 

\section{Functional derivative of the isolation strength $\merit$}\label{derivationAppendix}
In this appendix we describe the algebra that leads from  Eq.~\eqref{basic} to Eq.~\eqref{eq:dFNoFlip}, which essentially amounts to explicitly evaluating the two functional derivatives appearing Eq.~\eqref{basic}. Beginning with $\frac{\partial \merit}{\partial\G}(\rA,\rD)$, we have;
\begin{align}
    &\frac{\partial \merit}{\partial\G}(\rA,\rD) = \frac{\partial }{\partial\G(\rA,\rD) }\left[\frac{\left|\dA\cdot \G(\rA,\rD)\cdot \dD\right|^2}{\left|\dD\cdot \G(\rD,\rA)\cdot \dA\right|^2} \right] \notag \\
        &=\frac{1}{\left|\dD\cdot \G(\rD,\rA)\cdot \dA\right|^2} \frac{\partial }{\partial\G(\rA,\rD) }{\left|\dA\cdot \G(\rA,\rD)\cdot \dD\right|^2}    
\end{align}
Treating $\G$ and its conjugate as independent \cite{QEDBennett_2020}, we have;
\begin{align}
    \frac{\partial }{\partial\G(\rA,\rD) }&{\left|\dA\cdot \G(\rA,\rD)\cdot \dD\right|^2}\notag \\
    &= (\dA \otimes \dD) \dA\cdot \G^*(\rA,\rD)\cdot \dD
\end{align}
where $\otimes$ denotes the outer product so that $(\dA \otimes \dD)_{ij} = {\dA}_i {\dD}_j$. Thus,
\begin{align}
    &\frac{\partial \merit}{\partial\G}(\rA,\rD)=(\dA \otimes \dD) \frac{\dA\cdot \G^*(\rA,\rD)\cdot \dD}{\left|\dD\cdot \G(\rD,\rA)\cdot \dA\right|^2}.
\end{align}
Substituting this into Eq.~\eqref{basic} in the main text and using the following relation for arbitrary vectors $\mathbf{a}$, $\mathbf{b}$ and an arbitrary matrix $\mathbb{C}$
\begin{equation}\label{usefulVectorProperty}
    (\mathbf{a} \otimes \mathbf{b}) \odot \mathbb{C} = \mathbf{a} \cdot \mathbb{C} \cdot \mathbf{b},
\end{equation}
one finds the first term of Eq.~\eqref{eq:dFNoFlip}.

The derivative $\frac{\partial \merit}{\partial\G}(\rD,\rA)$ is more complicated as the relevant Green's tensor appears in the denominator. We have:
\begin{align}
    &\frac{\partial \merit}{\partial\G}(\rD,\rA) = \frac{\partial }{\partial\G(\rD,\rA) }\left[\frac{\left|\dA\cdot \G(\rA,\rD)\cdot \dD\right|^2}{\left|\dD\cdot \G(\rD,\rA)\cdot \dA\right|^2} \right] \notag    \\
    & =\left|\dA\cdot \G(\rA,\rD)\cdot \dD\right|^2 \frac{\partial }{\partial\G(\rD,\rA) }\left[\frac{1}{\left|\dD\cdot \G(\rD,\rA)\cdot \dA\right|^2} \right] \notag \\ 
        & =\frac{\left|\dA\cdot \G(\rA,\rD)\cdot \dD\right|^2}{\dD\cdot \G^*(\rD,\rA)\cdot \dA} \frac{\partial }{\partial\G(\rD,\rA) }\left[\frac{1}{\dD\cdot \G(\rD,\rA)\cdot \dA} \right] \label{dRdGDAStep} 
\end{align}
We therefore require a derivative of the form;
\begin{align}
    \frac{\partial}{\partial \mathbb{C}} \left(\frac{1}{\mathbf{a}\cdot \mathbb{C}\cdot  \mathbf{b}}\right)
\end{align}
where, again, $\mathbf{a}$ and $\mathbf{b}$ are arbitrary vectors and $\mathbb{C}$ is an abitrary matrix. Switching to index notation;
\begin{align}
    \left[\frac{\partial}{\partial \mathbb{C}} \left(\frac{1}{\mathbf{a}\cdot \mathbb{C}\cdot  \mathbf{b}}\right)\right]_{ij} = \frac{\partial}{\partial {C}_{ij}} \left(\frac{1}{\sum_{p,q}{a}_p {C}_{pq} {b}_q}\right)
\end{align}
Generalising the quotient rule from elementary calculus, we have;
\begin{align}
    \left[\frac{\partial}{\partial \mathbb{C}} \left(\frac{1}{\mathbf{a}\cdot \mathbb{C}\cdot  \mathbf{b}}\right)\right]_{ij} = -\frac{\left(\sum_{p,q}\delta_{ip}\delta_{jq} a_p b_q\right)}{\left(\sum_{s,t}{a}_s {C}_{st} {b}_t\right)^2}\notag \\
    =-\frac{ a_i c_j}{\left(\sum_{s,t}{a}_s {C}_{st} {b}_t\right)^2}
\end{align}
Therefore
\begin{align}
        \frac{\partial}{\partial \mathbb{C}} \left(\frac{1}{\mathbf{a}\cdot \mathbb{C}\cdot  \mathbf{b}}\right) =- \frac{\mathbf{a} \otimes \mathbf{b} }{(\mathbf{a}\cdot \mathbb{C}\cdot  \mathbf{b})^2},
\end{align}
which in turn means that
\begin{align}
    \frac{\partial }{\partial\G(\rD,\rA) }&\left[\frac{1}{\dD\cdot \G(\rD,\rA)\cdot \dA} \right] \notag \\
    &= -\frac{\dD \otimes \dA}{\left[\dD\cdot \G(\rD,\rA)\cdot \dA\right]^2 },
\end{align}
which can in turn be used in Eq.~\eqref{dRdGDAStep}
\begin{align}
    &\frac{\partial \merit}{\partial\G}(\rD,\rA) =\notag \\
    &-\frac{\left|\dA\cdot \G(\rA,\rD)\cdot \dD\right|^2}{\dD\cdot \G^*(\rD,\rA)\cdot \dA} \frac{\dD \otimes \dA}{\left[\dD\cdot \G(\rD,\rA)\cdot \dA\right]^2 }.
\end{align}
Substituting this into Eq.~\eqref{basic} in the main text and using Eq.~\eqref{usefulVectorProperty} yields the second term of Eq.~\eqref{eq:dFNoFlip}, thereby completing the explicit evaluation of the functional derivatives therein.


\providecommand{\noopsort}[1]{}\providecommand{\singleletter}[1]{#1}%

\end{document}